\documentclass[aps,twocolumn,prl,showpacs,floatfix,superscriptaddress]{revtex4}

\usepackage{amsfonts}
\usepackage{graphicx}
\usepackage{amsmath}
\usepackage{amssymb}

\input{tcilatex}

\begin{document}

\title{Liquid-Crystal State of $\nu=1/m$ Quantum Hall Effects}
\author{O. G. Balev}
\email[Electronic address: ]{vbalev@df.ufscar.br}
\affiliation{Departamento de F\'{\i}sica, Universidade Federal de S\~{a}o
Carlos, 13565-905, S\~{a}o Carlos, S\~{a}o Paulo, Brazil}
\affiliation{Institute of Semiconductor Physics, NAS of Ukraine,
Kiev 03650, Ukraine }

\begin{abstract}
At filling factor $\nu=1/m$, $m$ odd integer, I present variational
ground-state and excited-state wave functions, of
two-dimensional electron system with homogeneous
ion background, that show the condensation into a liquid-crystal
state. For $m=1, 3, 5$, the ground-state energy per electron
is substantially lower than the Laughlin one, for
uniform liquid state.
%
\end{abstract}

\date{\today}
\maketitle


Discovery of the fractional quantum Hall effect \cite{tsui1982} in
two-dimensional electron systems (2DES) of GaAs-based samples and the
Laughlin seminal theory of this effect \cite{laughlin1983} have generated
strong interest to properties of fractional quantum Hall states at $\nu=1/m$%
, especially for $m=3$ and $5$, \cite{prang1990,
chakraborty1995,sarma1997,yoshioka1983,
maki1983,yoshioka83i84,haldane1985,kivelson1986,morf1986,macdonald1986,
claro1987,willett1988,jane1989,
dorozhkin1995,mikhailov2001,wexler2004,
cabo2004}. Current understanding is that
for $m=3,\; 5$ the Laughlin wave function
\cite{laughlin1983} gives the best known analytical approximation of exact
many-body ground-state wave function
\cite{prang1990,
chakraborty1995,sarma1997,yoshioka1983,maki1983,yoshioka83i84,
haldane1985,kivelson1986,morf1986,macdonald1986,
claro1987,willett1988,jane1989,dorozhkin1995,
mikhailov2001,wexler2004,cabo2004}.
For $m=1$
the Laughlin wave function coincides with the Hartree-Fock approximation
(HFA) one \cite{laughlin1983,mikhailov2001}, built from the symmetric gauge
single-electron wave functions of the lowest Landau level, and leads to the
total energy per electron \cite{laughlin1983} $\epsilon_{HF}=-\sqrt{\pi/8}
e^{2}/(\varepsilon \ell_{0})$;
%
%
%
$\ell_{0}=\sqrt{\hbar c/|e|B}$ is the
magnetic length and $\varepsilon$ the background dielectric constant.
%
%
In present study strong many-body effects are essentially related as with
$N$ electrons of 2DES so with $N$ ions. I treat the ions on more
equal footing with 2DES than previously
\cite{laughlin1983,prang1990,
chakraborty1995,sarma1997,yoshioka1983,
maki1983,yoshioka83i84,haldane1985,kivelson1986,morf1986,macdonald1986,
claro1987,
%
mikhailov2001,wexler2004,cabo2004}. In addition,
I use more adequate sets of single-body wave functions; they
are localized mainly (or exactly) within the
%
%
unit cell
$L_{x}^{\square} \times L_{x}^{\square}$,
$(L_{x}^{\square})^{2}=L_{x} L_{y}/N$. These
%
%
wave functions help better reflect the
tendency: i) of an ion to be mainly localized within its own unit
cell, and ii) of an electron to be present mainly within any such
unit cell, with equal probability.

In this Letter, at filling factors $\nu=1/m$ with odd integer $m$, I present
many-body variational ground-state and excited-state wave functions for
electron-ion system, with homogeneous ion density, that have strong
correlations between 2DES and ions. The former wave function result in: i)
substantially lower ground-state energies for $\nu=1,\; 1/3, \; 1/5$ than
obtained in Ref. \cite{laughlin1983}; ii) the electron density,
Eq. (\ref{38d}), periodic along one direction with period
$\sqrt{2\pi/m} \; \ell_{0}$ is typically very weakly modulated;
iii) fractionally quantized Hall conductance, for $m=3,\; 5,\ldots$. I
obtain finite excitation gaps
%
%
along with fractional quasielectron, $e/m$, and quasihole, $|e|/m$, charges,
for $m \geq 3$.

We consider a zero-thickness 2DES of width $L_{y}$ ($L_{y}/2> y >-L_{y}/2$)
and of length $L_{x}$ ($L_{x}> x >0$) in the presence of a
%
%
magnetic
field, $\mathbf{B}= B \hat{\mathbf{z}}$. The Landau gauge for the vector
potential, $\mathbf{A}(\mathbf{r})=(-By,0,0)$, is used; $N$ electrons of a
2DES and $N$ ions
%
%
are located in the main region, $L_{x} \times L_{y}$.
%
%
%
%
As ions are very heavy, their kinetic energy can be neglected
\cite{laughlin1983,mikhailov2001}. Then the many-body Hamiltonian
%
%
$\hat{H}=\hat{H}_{0}+V_{ee}+V_{ei}+V_{ii}$ ,
%
%
where the kinetic energy of electrons $\hat{H}_{0}=\sum_{i=1}^{N} \hat{h}%
_{0i}$, $\hat{h}_{0j}=[i\hbar \mathbf{\nabla}_{j}+ e\mathbf{A}(\mathbf{r}%
_{j})/c]^{2}/2m^{\ast}$; the electron-electron potential
$V_{ee}=\frac{1}{2}\sum_{i=1}^{N} \sum_{j=1, j \neq
i}^{N} V(\mathbf{r}_{i}-\mathbf{r}_{j})$,
$\mathbf{r}_{i}=(x_{i},y_{i})$;
%
%
the electron-ion potential
$V_{ei}=-\sum_{i=1}^{N} \sum_{j=1}^{N}
V(\mathbf{r}_{i}-\mathbf{R}_{j})$;
%
%
and the ion-ion potential $V_{ii}=\frac{1}{2}\sum_{i=1}^{N} \sum_{j=1,
j \neq i}^{N} V(\mathbf{R}_{i}-\mathbf{R}_{j})$,
$\mathbf{R}_{i}=(X_{i},Y_{i})$;
%
%
$V(\mathbf{r})=e^{2}/\varepsilon |\mathbf{r}|$.
%
%

Let us for a $x-$stripe region, given by
$x \in(L_{x}^{\square}(n_{xs}^{\alpha}-1); L_{x}^{\square} n_{xs}^{\alpha})$
and $y \in(-\infty,\infty)$, introduce normalized solutions of
the single-electron Schr\"{o}dinger ($\omega_{c}=|e|B/m^{\ast}
c$, $k_{x\alpha}=2\pi n_{ys}^{\alpha}/L_{x}^{\square}$) equation
\begin{equation}
\hat{h}_{0} \psi_{n_{\alpha};n_{xs}^{\alpha}, k_{x\alpha}}^{L_{x}^{\square}}(%
\mathbf{r})= \hbar \omega_{c} (n_{\alpha}+1/2)
\psi_{n_{\alpha};n_{xs}^{\alpha},k_{x\alpha}}^{L_{x}^{\square}}(\mathbf{r})
\label{8d}
\end{equation}%
of ($\Psi_{n}(y)$ is the harmonic oscillator function) the form
\begin{equation}
\psi_{n_{\alpha};n_{xs}^{\alpha},k_{x\alpha}}^{L_{x}^{\square}}(\mathbf{r})
= \frac{e^{ik_{x\alpha} x}}{(L_{x}^{\square})^{1/2}} \Psi_{n_{%
\alpha}}(y-y_{0}(k_{x\alpha}))  ,  \label{9d}
\end{equation}%
where $y_{0}(k_{x\alpha})=\ell_{0}^{2} k_{x\alpha}$, $n_{ys}^{\alpha}=0,\pm
1,\dots, \pm(n_{ys}^{\max,t}-1)/2$; $n_{ys}^{\max,t}$ is the odd integer
such that $(2\pi/L_{x}^{\square}) n_{ys}^{\max,t} \ell_{0}^{2}=L_{y}$ .
%
%
%
%
For $x > L_{x}^{\square} n_{xs}^{\alpha}$ or
$x <L_{x}^{\square}(n_{xs}^{\alpha}-1)$,
$\psi_{n_{\alpha};n_{xs}^{\alpha},k_{x\alpha}}^{L_{x}^{\square}}(\mathbf{r})
\equiv 0$.
Here $n_{xs}^{\alpha}=1, 2,\ldots, n_{xs}^{\max}$ gives the number to
the $x-$stripe region and $L_{x}^{\square} n_{xs}^{\max} =L_{x}$.
Then the total number of states of the wave functions Eq. (\ref{9d}),
on the $n_{\alpha}-$th Landau level in the main region,
%
%
is $n_{xs}^{\max} n_{ys}^{\max,t} =L_{x}L_{y}/(2\pi
\ell_{0}^{2})=N_{L}$, which is equal to the number of states of ``usual''
wave functions \cite{landau1965} $\psi_{n_{\alpha};1,k_{x\alpha}}^{L_{x}}(%
\mathbf{r})$. Wave functions Eq. (\ref{9d}) are orthonormal as
\begin{eqnarray}
\int_{0}^{L_{x}} dx \int_{-\infty}^{\infty} &&dy \;
\psi_{n_{\beta};n_{xs}^{\beta},k_{x\beta}}^{L_{x}^{\square} \ast}(\mathbf{r}%
) \psi_{n_{\alpha};n_{xs}^{\alpha},k_{x\alpha}}^{L_{x}^{\square}}(\mathbf{r})
\notag \\
&&= \delta_{n_{\beta},n_{\alpha}} \delta_{n_{xs}^{\beta},n_{xs}^{\alpha}}
\delta_{k_{x\beta},k_{x\alpha}} .  \label{13d}
\end{eqnarray}%
It can be shown that the set of single-electron wave functions
Eq. (\ref{9d}) is complete. Eq. (\ref{13d}) reduces to well known
result \cite{landau1965} for
$\psi_{n_{\alpha};1,k_{x\alpha}}^{L_{x}}(\mathbf{r})$ if to change $%
L_{x}^{\square}$ on $L_{x}$.
Further, $(L_{x}^{\square})^{2}=L_{x} L_{y}/N=L_{x} L_{y}/\nu N_{L}$,
where $N$ is fixed for a given sample; hence, $L_{x}^{\square}$ is
also fixed. Then we obtain
\begin{equation}
L_{x}^{\square}/\Delta y_{0}=1/\nu .  \label{15d}
\end{equation}%
From (\ref{15d}) it is seen that within each unit cell can appear only an
odd integer number, $m=1, 3,\ldots$, of quantized oscillator centres, $%
y_{0}(k_{x\alpha})$;
%
%
%
even $m$, not treated here, is a special case.
Then Eq. (\ref{15d}) gives, $\ell=0,\; 1,\ldots$, that
\begin{equation}
1/\nu=m ,  \label{16d}
\end{equation}%
where $m=2\ell+1$. From Eqs. (\ref{15d}), (\ref{16d}) it follows
\begin{equation}
L_{x}^{\square }=\sqrt{2\pi m}\;\ell _{0}.  \label{18d}
\end{equation}%
As for $\nu=1/m$ there are $m$ quantized values of $y_{0}(k_{x\alpha i})$
within an $i-$th unit cell and each of them has a particular position within
the unit cell, we separate all $N_{L}$ states Eq. (\ref{9d}), of a $%
n_{\alpha}-$th Landau level, into the $m$ sets of wave functions. Within any
such $n-$th set of states $[y_{0}(k_{x\alpha j}^{(n)})- y_{0}(k_{x\alpha
i}^{(n)})]=k L_{x}^{\square}$, where $k$ is an integer. Here $j$($i$) is the
number of a unit cell; it can be any integer from $1$ to $N$. This $i-$%
number exactly defines the $i-$th unit cell. The superscript in $k_{x\alpha
i}^{(n)}$ is given to distinguish the $k_{x\alpha i}$ pertinent to the $n-$%
th set of states; the subscript (superscript), $i$, in $k_{x\alpha i}$, etc.
($n_{xs}^{(i)}$, etc.) indicates belonging to the $i-$th unit cell. We
choose the values of $n$ as $n=0,\ldots,\pm \ell$ and define $k_{x i}^{(n)}
\equiv k_{x\alpha i}^{(n)}$ as 
\begin{equation}
k_{x i}^{(0)}=(2\pi m/L_{x}^{\square}) n_{ys}^{(i)},\ldots, k_{x i}^{(\pm
\ell)}=k_{x i}^{(0)} \pm 2\pi \ell/L_{x}^{\square} ,  \label{19d}
\end{equation}%
where $n_{ys}^{(i)}=0, \pm 1,\ldots, \pm (n_{ys}^{\max}-1)/2$; $%
n_{ys}^{\max}=n_{ys}^{\max,t}/m$ is an odd integer. Wave functions Eq. (\ref%
{9d}) of the $n_{\alpha }=0$ Landau level we denote, at $\nu =1/m$, as well
as 
\begin{equation}
\varphi _{n_{xs}^{(i)},k_{xi}^{(n)}}^{i,(m)}(\mathbf{r})\equiv \psi
_{0;n_{xs}^{(i)},k_{xi}^{(n)}}^{L_{x}^{\square}}(\mathbf{r}).  \label{20d}
\end{equation}

I assume the ground-state wave function of electron-ion system, $\Psi
_{N,N}^{(m)}(\mathbf{r}_{1},\ldots ,\mathbf{r}_{N};\mathbf{R}_{1},\ldots ,%
\mathbf{R}_{N})$, in the form ($C_{n} \equiv C_{n}(m)$)
\begin{equation}
\Psi _{N,N}^{(m)}=\left[ \sum_{n=-\ell }^{\ell }C_{n}\Psi _{N}^{n,(m)}\right]
\;\dprod\limits_{i=1}^{N}\phi _{n_{xs}^{(i)},n_{ys}^{(i)}}^{(i)}(\mathbf{R}%
_{i}),  \label{45d}
\end{equation}%
where $|C_{n}|^{2}=1/m$, and the \textquotedblleft
partial\textquotedblright\ many-electron wave function, $\Psi _{N}^{n,(m)}(%
\mathbf{r}_{1},\mathbf{r}_{2},\ldots ,\mathbf{r}_{N})$, is $N-$dimensional
Slater determinant of wave functions Eq. (\ref{20d}).
%
%
%
%
%
%
Here \textquotedblleft single-ion\textquotedblright\
wave functions $\phi _{n_{xs}^{(i)},n_{ys}^{(i)}}^{(i)}(\mathbf{R})$
%
%
are introduced as: $|\phi
_{n_{xs}^{(i)},n_{ys}^{(i)}}^{(i)}(\mathbf{R})|^{2}=1/(L_{x}^{\square })^{2}$%
, if both $X\in (L_{x}^{\square }(n_{xs}^{(i)}-1),L_{x}^{\square
}n_{xs}^{(i)})$ and $Y\in (L_{x}^{\square }(n_{ys}^{(i)}-1/2),L_{x}^{\square
}(n_{ys}^{(i)}+1/2))$; if $X$ or/and $Y$ is outside of this $i-$th unit cell
then $\phi _{n_{xs}^{(i)},n_{ys}^{(i)}}^{(i)}(\mathbf{R})\equiv 0$. The set
of these single-body wave functions is orthonormal; then, $\langle \Psi
_{N,N}^{(m)}|\Psi _{N,N}^{(m)}\rangle =1$. The electron density, $n(\mathbf{r%
})=\langle \Psi _{N,N}^{(m)}|\sum_{j=1}^{N}\delta (\mathbf{r}-\mathbf{r}%
_{j})|\Psi _{N,N}^{(m)}\rangle $, in the main region, for $n_{ys}^{\max
}\ggg 1$ (then $N\ggg 1$), is
\begin{equation}
n(y)=\frac{\ell _{0}^{-2}}{2\pi m}[1+2\sum_{k=1}^{\infty }e^{-\pi
mk^{2}/2}\cos (\frac{\sqrt{2\pi m}}{\ell _{0}}ky)],  \label{38d}
\end{equation}%
after using the Fourier transformations and the Poisson's summation
formula \cite{hilbert}. Eq. (\ref{38d}) gives that a unit cell is
\textquotedblleft dressed\textquotedblright\ by electron charge, $e$. The
ion density $n_{io}(\mathbf{r})=n_{io}$, where $n_{io}=1/(2\pi m\ell
_{0}^{2})$.
Point out, in a good approximation of experimental conditions,
$\Psi _{N,N}^{(m)}$ gives that each ion is located in its own unit cell.
%
%

The total energy in the ground-state Eq. (\ref{45d}) is
\begin{equation}
E_{N}^{(m)}=\langle \Psi _{N,N}^{(m)}|\hat{H} |\Psi _{N,N}^{(m)}\rangle ,
\label{80d}
\end{equation}%
where the kinetic energy term gives $m^{-1} \sum_{n=-\ell }^{\ell }\langle
\Psi _{N}^{n,(m)}|\hat{H}_{0}|\Psi_{N}^{n,(m)}\rangle =\hbar \omega_{c} N/2$%
, cf. with Ref. \cite{yoshioka83i84};
details will be published elsewhere \cite{balev2004}.
In Eq. (\ref{80d}) the term $\langle \Psi _{N,N}^{(m)}|V_{ii}|\Psi
_{N,N}^{(m)}\rangle$ obtains the form
\begin{eqnarray}
&&\frac{1}{2} \sum_{i=1}^{N} \sum_{j=1, j \neq i}^{N}
\int_{-\infty}^{\infty} d \mathbf{R} \int_{-\infty}^{\infty} d \mathbf{R}%
^{\prime} \frac{e^{2}}{\varepsilon |\mathbf{R}-\mathbf{R}^{\prime}|}  \notag
\\
&&\;\;\; \times |\phi_{n_{xs}^{(i)},n_{ys}^{(i)}}^{(i)}(\mathbf{R})|^{2} \;
\; |\phi_{n_{xs}^{(j)},n_{ys}^{(j)}}^{(j)}(\mathbf{R}^{\prime})|^{2} .
\label{82d}
\end{eqnarray}%
$\langle \Psi_{N,N}^{(m)}|V_{ei}|\Psi _{N,N}^{(m)}\rangle$ in Eq. (\ref{80d}%
) has the form
\begin{eqnarray}
&&-\; \frac{1}{m} \sum_{n=-\ell }^{\ell } \sum_{i=1}^{N} \sum_{j=1}^{N}
\int_{-\infty}^{\infty} d \mathbf{r} \int_{-\infty}^{\infty} d \mathbf{R}
\frac{e^{2}}{\varepsilon |\mathbf{r}-\mathbf{R}|}  \notag \\
&&\;\;\; \;\;\;\; \times |\varphi_{n_{xs}^{(i)},k_{xi}^{(n)}}^{i,(m)}(%
\mathbf{r})|^{2} \; |\phi_{n_{xs}^{(j)},n_{ys}^{(j)}}^{(j)}(\mathbf{R})|^{2}
.  \label{83d}
\end{eqnarray}%
$\langle \Psi_{N,N}^{(m),eh}|V_{ee}|\Psi _{N,N}^{(m),eh}\rangle$ in Eq. (\ref%
{80d}) we can rewrite as
\begin{eqnarray}
\frac{1}{2m} \sum_{n=-\ell}^{\ell} \sum_{i=1}^{N} \sum_{j=1,j \neq i}^{N}
\langle \Psi_{N}^{n,(m)}| \frac{e^{2}}{\varepsilon |\mathbf{r}_{i}-\mathbf{r}%
_{j}|} |\Psi _{N}^{n,(m)} \rangle ,  \label{54d}
\end{eqnarray}%
where the matrix elements are calculated as in HFA \cite{madelung1981}. Then
Eq. (\ref{80d}) is written, $\tilde{E}_{N}^{(m)}=E_{N}^{(m)}-\hbar
\omega_{c} N/2$, as
\begin{equation}
\tilde{E}_{N}^{(m)}= \frac{e^{2} N}{\varepsilon \ell_{0}}
\left[ F_{2}^{A}(m)+F_{1}^{C}(m)+\Delta F_{1}^{C}(m) \right] ,
\label{84d}
\end{equation}%
%
%
where $F_{2}^{A}(m)$ is exchange-alike term from Eq. (\ref{54d}),
\begin{eqnarray}
F_{2}^{A}(m) &=& -\frac{1}{\pi } \sum_{k=-\infty; k \neq 0}^{\infty }
e^{-\pi m\;k^{2}} \int_{0}^{\infty }d\xi \int_{0}^{\infty }d\eta  \notag \\
&& \times e^{-\eta ^{2}/2} \; S_{m}^{2}(\xi) \; G_{m}(\xi,\eta;k) ,
\label{65d}
\end{eqnarray}
$G_{m}(\xi,\eta;k)=[(\xi -\sqrt{2\pi m}\;k)^{2}+\eta^{2}]^{-1/2}$.
Further,
%
%
from the ``diagonal'' part, $i=j$, of Eq. (\ref{83d})
it follows the term
\begin{eqnarray}
F^{C}_{1}(m)&=&-\;\frac{2}{\pi} \int_{0}^{\infty} d \xi \int_{0}^{\infty} d
\eta e^{-\eta^{2}/4} \; (\xi^{2}+\eta^{2})^{-1/2}  \notag \\
&&\times f_{m}(\eta) \; S_{m}(\eta) \; S_{m}^{2}(\xi) ,  \label{87d}
\end{eqnarray}%
where $S_{m}(x)=\sin(\sqrt{\pi m/2}\; x)/(\sqrt{\pi m/2}\; x)$, $%
f_{1}(\eta)=1$; for $m=3, 5,\ldots$, $f_{m}(\eta)=m^{-1} \left[%
1+2\sum_{n=1}^{\ell} \cos(\sqrt{2\pi/m} \; n \; \eta ) \right]$.
%
%
The sum of i) ``nondiagonal'' part, $i
\neq j$, of Eq. (\ref{83d}), ii) Eq. (\ref{82d}), and iii) direct-alike
contribution from Eq. (\ref{54d}) gives
\begin{eqnarray}
\Delta F^{C}_{1}(m)&=&-\frac{1}{\pi} \int_{0}^{\infty} d \xi
\int_{0}^{\infty} d \eta \; g_{m}(\eta) S_{m}^{2}(\xi)/\sqrt{\xi^{2}+\eta^{2}%
}  \notag \\
&&+(1/\sqrt{2 \pi \; m}) \sum_{k=1}^{\infty} k^{-1} \; e^{-\pi \; k^{2}/m} ,
\label{89d}
\end{eqnarray}%
%
where
$g_{m}(\eta)=S_{m}^{2}(\eta)+e^{-\eta^{2}/2} -2 e^{-\eta^{2}/4} \;
f_{m}(\eta)\; S_{m}(\eta)$. We can rewrite Eq. (\ref{84d}) as $\tilde{E}%
_{N}^{(m)}/[e^{2} N/(\varepsilon \ell_{0})]=U^{C}(m)$, where $%
U^{C}(m)=[F^{C}_{1}(m)+\Delta F^{C}_{1}(m)+F_{2}^{A}(m)]$ gives lowering of
the total energy per electron in the units of $e^{2}/\varepsilon \ell_{0}$.
I calculate numerically that $U^{C}(1) \approx -1.202775$,
$U^{C}(3) \approx -0.712971$, $U^{C}(5) \approx -0.552704$, and $%
U^{C}(7) \approx -0.466528$. Here $U^{C}(1)$, $U^{C}(3)$, and $U^{C}(5)$ are
substantially lower than pertinent total lowering at $\nu=1, \;1/3$, and $1/5
$ for the Laughlin variational wave function \cite{laughlin1983} $\; -\;
\sqrt{\pi/8} \approx -0.6267$, $-0.4156 \pm 0.0012$, and $-0.3340 \pm 0.0028$%
, respectively.
%
%
Notice, for $m=1$, if in Eqs. (\ref{80d})-(\ref{54d}) formally to change
both single-electron and ``single-ion'' wave functions on ``usual''
single-particle wave function
%
%
$\psi_{0;1,k_{x\alpha}}^{L_{x}}$, then
$\tilde{E}_{N}^{(m)}=N \epsilon_{HF}$, for $L_{x} \rightarrow \infty$.

I assume, $\Psi _{N,N;(m)}^{i_{0};j_{0}}(\mathbf{r}_{1},\ldots ,\mathbf{r}%
_{N};\mathbf{R}_{1},\ldots ,\mathbf{R}_{N})$, excited-state wave function of
the ground-state Eq. (\ref{45d}) as
\begin{eqnarray}
&&\Psi _{N,N;(m)}^{i_{0};j_{0}}=\dprod\limits_{i=1}^{N}\phi
_{n_{xs}^{(i)},n_{ys}^{(i)}}^{(i)}(\mathbf{R}_{i})\sum_{n=-\ell }^{\ell }%
\tilde{C}_{n}[(1-\delta _{n,0})  \notag \\
&&\;\;\;\;\;\times \Psi _{N}^{n,(m)}(\mathbf{r}_{1},\ldots)+\delta
_{n,0}\Phi _{N;(m)}^{i_{0};j_{0}}(\mathbf{r}_{1},\ldots)],
\label{115d}
\end{eqnarray}%
where $\tilde{C}_{n}=C_{n}$, for $n\geq 0$, and $\tilde{C}_{n}=-\;C_{n}$,
for $n<0$. An excited \textquotedblleft partial\textquotedblright\
many-electron wave function $\Phi _{N;(m)}^{i_{0};j_{0}}$ it follows from
the $\Psi _{N}^{0,(m)}$
%
%
after changing of the $i_{0}-$th
row, $\varphi _{n_{xs}^{(i_{0})},k_{xi_{0}}^{(0)}}^{i_{0},(m)}(\mathbf{r}%
_{1}),\ldots ,\varphi _{n_{xs}^{(i_{0})},k_{xi_{0}}^{(0)}}^{i_{0},(m)}(%
\mathbf{r}_{N})$, by the determinant row of the, for $m \geq 3$, form $%
\varphi _{n_{xs}^{(j_{0})},k_{xj_{0}}^{(\tilde{n})}}^{j_{0},(m)}(\mathbf{r}%
_{1}),\ldots ,\varphi _{n_{xs}^{(j_{0})},k_{xj_{0}}^{(\tilde{n}%
)}}^{j_{0},(m)}(\mathbf{r}_{N})$, where $\tilde{n}\neq 0$; Eq. (\ref{19d})
gives $k_{xj_{0}}^{(\tilde{n})}=(2\pi \;m/L_{x}^{\square })\left[
n_{ys}^{(j_{0})}+\tilde{n}/m\right] $, $\tilde{n}=\pm 1,\ldots ,\pm \ell $;
i.e., $k_{xj_{0}}^{(\tilde{n})}\neq k_{xi}^{(0)}$, where $i=1,\ldots ,N$.
For $m=1$ in new determinant row: i) the implicit spin up wave function $%
|1>=\psi _{1}(\sigma _{j})$ should be substituted by spin down one, $%
|-1>=\psi _{-1}(\sigma _{j})$; ii) $\tilde{n}=0$.
We assume that the $i_{0}$-th unit cell,
%
%
where the quasihole appears, as well as the $j_{0}-$th unit cell, where
the quasielectron is mainly localized, there are well inside of the
main region. $\Psi_{N,N;(m)}^{i_{0};j_{0}}$ describes excitation of a
quasiexciton type \cite{laughlin1983,prang1990,
chakraborty1995,sarma1997,morf1986,macdonald1986}. It is seen that
at any separation between the quasielectron and the quasihole their
charges are given (details will be published elsewhere
\cite{balev2004}) as $e/m$ and $|e|/m$, respectively.
%
%
We need the energy gap, $\Delta ^{(m)}$, for the creation of one
quasielectron and one quasihole, infinitely spatially separated\cite%
{laughlin1983,prang1990,
chakraborty1995,sarma1997,morf1986,macdonald1986,willett1988,dorozhkin1995};
notice, $\langle \Psi _{N,N;(m)}^{i_{0};j_{0}}|\Psi
_{N,N}^{(m)}\rangle =0$ and $\langle \Psi _{N,N;(m)}^{i_{0};j_{0}}|\Psi
_{N,N;(m)}^{i_{0};j_{0}}\rangle =1$.

With infinitely spatially separated quasielectron and quasihole,
$\Delta ^{(m)}=\langle \Psi_{N,N;(m)}^{i_{0};j_{0}}|\hat{H}|
\Psi_{N,N;(m)}^{i_{0};j_{0}}\rangle -E_{N}^{(m)}$ is given,
for $m=3, 5, 7$, as
$\tilde{\Delta}^{(m)}=\left[|F_{1}^{C}(m)| +\Delta F^{C}(m,1) +
2 |F_{2}^{A}(m)|-F^{(m)}_{1} \right]/m$,
$\tilde{\Delta}^{(m)}=\Delta^{(m)}/(e^{2}/\varepsilon \ell_{0})$,
where
%
%
%
\begin{eqnarray}
F^{(m)}_{\tilde{n}}&=&(1/\pi) \sum_{k=-\infty}^{\infty }e^{-\pi m\;(k-\tilde{%
n}/m)^{2}}\int_{-\infty}^{\infty }d\xi \int_{0}^{\infty }d\eta  \notag \\
&&\times e^{-\eta ^{2}/2} \; S_{m}^{2}(\xi) \; G_{m}(\xi,\eta;k-\tilde{n}/m)
,  \label{A.4}
\end{eqnarray}
\begin{eqnarray}
&&\Delta F^{C}(m,\tilde{n})=-\; \frac{2^{3/2}}{\sqrt{\pi} m^{3/2}}
\sum_{k=1}^{\infty} \frac{\exp(-\frac{\pi k^{2}}{m})}{k} \sin^{2}\left(\frac{%
\pi k \tilde{n}}{m}\right)  \notag \\
&&+\frac{2}{\pi m} \int_{0}^{\infty} \int_{0}^{\infty} \frac{ d \xi d \eta}{%
\sqrt{\xi^{2}+\eta^{2}}} \; S_{m}^{2}(\xi) \; G_{m}(\eta) ,  \label{123d}
\end{eqnarray}%
$G_{m}(\eta)=\exp(-\eta^{2}/4) \; [ \exp(-\eta^{2}/4) -S_{m}(\eta) ]$, and
it is taken into account that only $\tilde{n}=1$ corresponds to
$\Delta^{(m)}$.
%
%
For $m=1$, $\Delta^{(1)}-|g_{0}| \mu_{B}
B=(e^{2}/\varepsilon \ell_{0}) [|F_{1}^{C}(1)| +\Delta F^{C}(1,0)+2
|F_{2}^{A}(1)|]$, where $g_{0}$ is the bare Land\'{e} g-factor.
%
%
I calculate
numerically that $\tilde{\Delta}^{(1)}- |g_{0}| \mu_{B} B/(e^{2}/\varepsilon
\ell_{0}) \approx 1.253895$ (i.e., very close to $\sqrt{\pi/2} \approx
1.253314$), $\tilde{\Delta}^{(3)} \approx 0.170657$, $\tilde{\Delta}%
^{(5)} \approx 0.069867$, and $\tilde{\Delta}^{(7)} \approx 0.036086$.

Point out that the ground-state Eq. (\ref{45d}) shows broken symmetry
``liquid-crystal'' behavior of 2DES as the electron density, Eq. (\ref{38d}%
), is periodic along $y$-direction, with period $\ell_{0} \sqrt{2\pi/m}$.
%
We can make electron density much more homogeneous,
%
%
however, the latter state has
much higher energy than $U^{C}(m)$.

For the ground-state Eq. (\ref{45d}), at $\nu=1/m$, I calculate (details
will be published elsewhere \cite{balev2004}) that the Hall conductance
%
%
%
%
%
$\sigma_{H}=e^{2}/( 2m \pi \hbar)$; i.e., it is properly quantized.
%
%
Similar to Refs. \cite{laughlin1983,niu1985}, we
can speculate that for a weak disorder
%
%
if the Fermi level still lies in a gap or mobility
gap the Hall conductance should be quantized in a finite range of $B$.
%
%

Present energy gap $\tilde{\Delta}^{(3)}$
%
%
is about $1.6$ times
larger than typically calculated for the Laughlin liquid pertinent
excitation gap \cite{haldane1985,morf1986,macdonald1986}.
%
%
%
For detailed comparison of the gap
with experiment it is known that a finite thickness of 2DES should be taken
into account as well as effects of disorder \cite{willett1988,morf2002}.
%
%
In addition, we can speculate that many-body
effects similar to those studied in \cite{balev2001}
%
%
(for \textquotedblleft traditional\textquotedblright\ $\nu =1$ state)
and related with edge states here, maybe, also will lead to highly
asymmetric pinning of the Fermi level within the energy gap.
%
%
Then, similar to \cite{balev2001}, actual activation gap can
be much smaller than $\tilde{\Delta}^{(m)}/2$.

In summary, I have presented, at $\nu=1/m$,
the theory
of liquid-crystal ground-state
with periodic, along one direction, density of 2DES and uniform
density of ions. The ground-state has strong correlations between
2DES and ions. The Hall conductance is properly quantized.
Excitation gap, for $m=1, \; 3, \; 5, \; 7$, is finite;
quasielectron and quasihole charges are
fractional, $\pm e/m$, for $m \geq 3$.
%

\acknowledgements

This work was supported in part by Universidade Federal de S\~{a}o Carlos.

\end{document}